\documentclass{PoS}

\usepackage{amsmath}

\newcommand{\cN}{{\cal N}}
\newcommand{\cD}{{\cal D}}
\newcommand{\cP}{{\cal P}}

\newcommand{\e}{\text{e}}
\newcommand{\ii}{\mathrm{i}}
\DeclareMathOperator{\Det}{Det}
\DeclareMathOperator{\Tr}{Tr}
\DeclareMathOperator{\tr}{tr}

\newcommand{\rk}{\right)}
\newcommand{\lk}{\left(}

\newcommand{\bra}[1]{\langle #1\rvert}
\newcommand{\ket}[1]{\lvert#1\rangle}

\newcommand{\be}{\begin{equation}}
\newcommand{\ee}{\end{equation}}

\newcommand{\bal}{\begin{align}}
\newcommand{\eal}{\end{align}}
\newcommand{\bea}{\begin{eqnarray}}
\newcommand{\eea}{\end{eqnarray}}

\newcommand*{\vev}[1]{\left< #1 \right>}
\providecommand*{\coloneq}{\mathrel{\mathop:}=}
\newcommand*{\dd}[1][]{\mathop{\mathrm{d}^{#1}}\mkern-4mu}

\newcommand{\vA}{\vec{A}}
\newcommand{\vx}{\vec{x}}
\newcommand{\vp}{\vec{p}}
\newcommand{\va}{\vec{a}}
\newcommand{\vd}{\vec{d}}
\newcommand{\vB}{\vec{B}}
\newcommand{\vD}{\vec{D}}
\newcommand{\vy}{\vec{y}}
\newcommand{\ve}{\vec{e}}

\title{Hamiltonian Approach to QCD: The effective potential of the Polyakov loop}

\ShortTitle{Hamiltonian Approach to QCD: The effective potential of the Polyakov loop}

\author{\speaker{Hugo Reinhardt}\\
Universit\"at T\"ubingen, Institut f\"ur Theoretische Physik\\
Auf der Morgenstelle 14, 72076 T\"ubingen, Germany\\
E-mail: \email{hugo.reinhardt@uni-tuebingen.de}}

\author{Jan Heffner\\
Universit\"at T\"ubingen, Institut f\"ur Theoretische Physik\\
Auf der Morgenstelle 14, 72076 T\"ubingen, Germany
}
\abstract{
The effective potential of the order parameter for confinement is calculated within the Hamiltonian approach
to Yang--Mills theory. Compactifying one spatial dimension and using a background gauge fixing
this potential is obtained by minimizing the energy density for a given background field. 
Using Gaussian type trial wave functionals I establish an analytic relation between the propagators in the
background gauge at finite temperature and the corresponding zero temperature propagators in 
Coulomb gauge. In the simplest truncation, neglecting the ghost and using the ultraviolet form of the 
gluon energy one recovers the Weiss potential. From the fully non-perturbative potential (with the ghost
included) one extracts a critical temperature of the deconfinement phase transition of 270 MeV for the 
gauge group SU$(2)$.}

\FullConference{Xth Quark Confinement and the Hadron Spectrum,\\
		October 8-12, 2012\\
		TUM Campus Garching, Munich, Germany}

\begin{document}
\section{Introduction}\label{sectionI}
Understanding the deconfinement phase transition is one of the major challenges of particle physics. In quenched QCD reliable results are obtained within the lattice approach. This approach fails, however, at large baryon density due to the notorious fermion sign problem. Therefore alternative non-perturbative approaches to continuum QCD are desirable. In recent years a variational approach to Yang--Mills theory in Coulomb gauge was developed \cite{FeuRei04}, which has provided a decent description of the infrared sector of the theory \cite{EppReiSch07, SchLedRei06, Campagnari:2008yg,ReiEpp07,Pak:2009em,Reinhardt:2008ek}. Recently this approach was extended to finite temperature \cite{HefReiCam12} and also to full QCD \cite{Pak:2011wu}. In this talk I will report on the calculation of the effective potential of the confinement order parameter within the Hamiltonian approach \cite{Reinhardt:2012qe}.

In quantum field theory the temperature $T$ is most easily introduced by compactifying the Euclidean time and interpreting the length $L$
of the compactified time interval as inverse temperature. In finite temperature Yang-Mills theory the order parameter of confinement
is the expectation value of the Polyakov loop
\be
\label{1}
P [A_0] = \frac{1}{N} \tr \cP \e^ { - \int^L_0 d x^0 A_0 \lk x^0, \vx \rk} \, .
\ee
The quantity $\langle P [A_0] (\vx) \rangle \sim \exp \left[ - F_\infty (\vx) L \right]$ is related to the free energy of a (infinitely heavy)
quark at spatial position $\vx$. In the confined phase this quantity vanishes by center symmetry while it is non-zero in the deconfined
phase, where center symmetry is broken. In continuum Yang-Mills theory the Polyakov loop is most easily calculated in Polyakov gauge 
$\partial_0 A_0 = 0 \, , \, A_0 =$ diagonal. In the fundamental modular region $0 < A_0 L / 2 < \pi$ the Polyakov loop
$P [A_0]$ is a unique function of the field $A_0$, which, for SU$(2)$, is given by $P [A_0] = \cos \lk A_0 L / 2 \rk$. As a consequence
of this relation and of Jenssen's inequality one can use instead of $\langle P [A_0] \rangle$ alternatively $P  [\langle A_0 \rangle]$ or
$\langle A_0 \rangle$ as order parameter of confinement, Refs.~\cite{Marhauser:2008fz, Braun:2007bx}. 
The order parameter of confinement can be most easily obtained by calculating the effective potential $e [a_0]$ of a 
temporal background field $a_0$ chosen in the Polyakov gauge and by calculating the Polyakov line (\ref{1}) from 
the field configuration $\bar{a}_0$ which minimizes $e [a_0]$, i.e. $\langle P [A_0] \rangle \simeq P [\bar{a}_0]$. The
effective potential $e [a_0]$ was first calculated in Refs. \cite{Gross:1980br, Weiss:1980rj} in 1-loop perturbation theory and is shown in 
Fig.~\ref{fig1-1}. This potential is minimal for a vanishing field and the order parameter accordingly yields $P \left[\bar{a}_0 = 0 \right] = 1$, which indicates
the deconfining phase. The aim of the present work is to give a non-perturbative evaluation of $e [a_0]$ \cite{Reinhardt:2012qe} in the Hamilton approach to Yang-Mills
theory \cite{FeuRei04}.

It is obvious that the effective potential of $\langle A_0 \rangle$ cannot be straightforwardly evaluated in the Hamiltonian approach since the
letter assumes Weyl gauge $A_0 = 0$. However, we can exploit O$(4)$ invariance of Euclidean quantum field theory and compactify instead of the
time one spatial axis (for example the $x_3$-axis) to a circle and interpret the length $L$ of the compactified dimension as inverse temperature.
Therefore we will consider in the following Yang-Mills theory at a finite compactified length $L$ in a constant color diagonal background field
$a_3$ and calculate the effective potential $e [a_3]$. In the Hamiltonian approach the effective potential $e [\va]$ of a spatial background field $\va$ is given by the minimum of the 
energy density $\vev{H}/V$ calculated under the constraint $\langle \vA \rangle = \va$. This minimal property of the effective potential calls for a 
variational calculation.

\section{Hamilton approach in background gauge}\label{sectionII}

In the presence of an external constant background field $\va$ the Hamiltonian approach can be most conveniently formulated in the 
background gauge
\be
\label{4}
\left[ \vd, \vA \right] = 0 \, , \quad \vd = \vec{\partial} + \va \, 
\ee
where all fields are taken in the adjoint representation. This gauge allows for an explicit resolution of Gauss' law, which results in the 
gauge fixed Hamiltonian
\be
\label{5}
H = \frac{1}{2} \int \dd^3 x \lk J_A^{- 1} \vec{\Pi} (\vx) J_A \cdot \vec{\Pi} (\vx) + \vB^2 (\vx) \rk + H_\text{C} \, ,
\ee
where $\vec{\Pi} = - \ii \delta / \delta \vA$ is the momentum operator of the gauge fixed field and 
\be
\label{6}
J_A = \Det \lk - {\vD} \cdot {\vd} \rk \,, \quad {\vD} = \vec{\partial} + {\vA}
\ee
is the Faddeev-Popov determinant. Furthermore,
\be
\label{7}
H_\text{C} = \frac{g^2}{2} \int \dd^3 x \, \dd^3 y\,J_A^{- 1} \, \rho^a (\vx) J_A \, F^{ab} (\vx, \vy) \rho^b (\vy)
\ee
is the analogue of the so-called Coulomb term which results from the kinetic term of the ``longitudinal'' part of the momentum operator. Here
\be
\label{8}
\rho^a = - {\vD} \cdot \vec{\Pi} = - \lk {\vA} - {\va} \rk \cdot \vec{\Pi}
\ee
is the color charge density of the gluons, which interacts through the kernel
\be
\label{9}
F = \lk - {\vD} \cdot {\vd} \rk^{- 1} \lk - {\vd} \cdot {\vd} \rk \lk - {\vD} \cdot {\vd} \rk^{- 1} 
\, .
\ee
For a vanishing background field $\va = 0$ the gauge (\ref{4}) reduces to the ordinary Coulomb gauge and the Hamiltonian $H$ (\ref{5})
becomes the familiar Yang-Mills Hamiltonian in Coulomb gauge, \cite{ChrLee}. Furthermore, as was shown in Ref.~\cite{HefReiCam12} the Coulomb term $H_\text{C}$ (\ref{7}) is negligible in the
gluon sector. Therefore we will ignore this term in the following.

We are interested here in the energy density in the state $\psi_a [A]$ minimizing
\be
\vev{ H }_a \coloneq \bra{\psi_a } H \ket{\psi_a }
\ee
under the constraint $\langle \vA \rangle_a = \va$. For this purpose we perform a variational calculation with the trial
wave functional 
\be
\label{10}
\psi_a [A]  =  J_A^{- 1/2} \tilde{\psi} [A - a]\,,\quad
\tilde{\psi} [A]  = \cN \e^{\, - \frac{1}{2} \int A \omega 
A } \, ,
\ee
which already fulfills the constraint $\langle \vA \rangle_a = \va$. For $\va = 0$ this ansatz reduces to the trial wave functional used
in Coulomb gauge \cite{FeuRei04}.
Proceeding as in the variational approach in Coulomb gauge \cite{FeuRei04}, from 
$\langle H \rangle_a \to min$ one derives a set of coupled equations for the gluon and ghost propagators
\begin{align}
\label{11}
\cD &= \langle A A  \rangle_{a=0} = \frac{1}{2} \omega ^{-1}\,,\quad G =  
- \vev{ \lk  ( \hat{\vec{D}} + \hat{\va} ) \hat{\vd} \rk^{- 1} }_{a=0} \, .
\end{align}
Using the same approximation as in Ref.~\cite{HefReiCam12} in Coulomb gauge, i.e. restricting to two loops in the energy, while neglecting $H_\text{C}$ (\ref{7}) and also the tadpole arising from the non-Abelian part of the magnetic energy, one finds from the
minimization of $\langle H \rangle_a$ the gap equation 
\be
\label{12}
\omega^2 =   - {\vd} \cdot {\vd}  + \chi^2\,,
\ee
where\footnote{We use here the compact notation $A (1) \equiv A^{a_1}_{i_1} (\vx_1)$. For Lorentz scalars like the ghost, the index ``$1$'' stands for 
the color index $a_1$ and the spatial position $\vx_1$.~Repeated indices are summed/integrated over.}
\be
\label{13}
\chi (1, 2) = - \frac{1}{2} \vev{ \frac{\delta^2 \ln J [A + a]}{\delta A (1) \delta A (2)} }_{a=0} 
= \frac{1}{2} \Tr \left[ G \Gamma (1) G \Gamma_0 (2) \right]
\ee
is the ghost loop (referred to as ``curvature'') with $\Gamma_0$ and $\Gamma$ being the bare and full ghost-gluon vertex. 
The gap equation (\ref{12}) has to be solved together with the Dyson-Schwinger equation (DSE) for the ghost propagator
\be
\label{14}
G^{- 1} = - {\vd} \cdot {\vd} - \Gamma_0 (1) G \Gamma (2) \cD (2, 1)\,.
\ee
Due to the presence of the background field these equations have a non-trivial color structure. Fortunately, due to the choice
of the background gauge (\ref{4}), the background field enters these equations only in form of the covariant derivative ${d} = \partial + {a}$.

We are interested in the effective potential of a background field in the Cartan algebra, which for SU$(2)$ has the form $\va T_3$.
This field becomes diagonal in the Cartan basis defined by the eigenvectors of the generators of the Cartan subgroup $T_3 |
\sigma \rangle = - \ii \sigma | \sigma \rangle \, , \, \sigma = 0, \pm 1$. By analyzing our equations of motion, i.e. the gap equation and 
the ghost Dyson-Schwinger equation, one finds that these equations have solutions, which are diagonal in the Cartan basis
\be
\label{15}
\cD^{\sigma \tau} (\vp) = \delta^{\sigma \tau} \cD^\sigma (\vp) \,, \quad G^{\sigma \tau} (\vp) = \delta^{\sigma \tau} G^\sigma (\vp) \, .
\ee
This is not surprising since the source of the non-trivial color structure is the background field and if this field is chosen 
to be diagonal in color space the same should be true for all propagators. In addition, one can show that the propagators in the 
presence of the background field in background gauge are related to the propagators in Coulomb gauge (i.e. in the absence of the
background field), $\cD (\vp)$, $G (\vp)$, by 
\be
\label{18}
\cD^\sigma (\vp) = \cD (\vp^\sigma) \,, \quad G^\sigma (\vp) = G (\vp^\sigma) \, ,
\ee
where 
\be
\label{19}
\vp^\sigma = \vp - \sigma \va
\ee
is the momentum shifted by the background field. In this way the results of the variational calculation in Coulomb gauge (in the absence of 
the background field) are sufficient to determine the propagators in the presence of the background field in background gauge.

Lattice calculation \cite{BurQuaRei09} of the gluon propagator in Coulomb gauge show that the gluon energy can be nicely fitted by Gribov's formula \cite{Gribov78}
\be
\label{22}
\omega (\vp) = \sqrt{\vp^2 + M^4/\vp^2} \, .
\ee
A full self-consistent solution of the gap equation (\ref{12}) and the ghost DSE (\ref{14}) reveals that $\omega (\vp)$ contains in addition sub-leading UV-logs,
which on the lattice are found to be small. Using Gribov's formula (\ref{22}) for $\omega (\vp)$ and solving the gap equation (\ref{12})
for $\chi (\vp)$ yields 
\be
\label{23}
\chi (\vp) = M^2 / |\vp| \, ,
\ee
which is indeed the correct IR-behavior obtained in a full solution \cite{HefReiCam12} of the coupled ghost DSE and gap equation but which misses the sub-leading
UV-logs.
\section{The effective potential}
\begin{figure}[t]
\centering
\parbox[t]{.45\linewidth}{
\includegraphics[width=\linewidth]{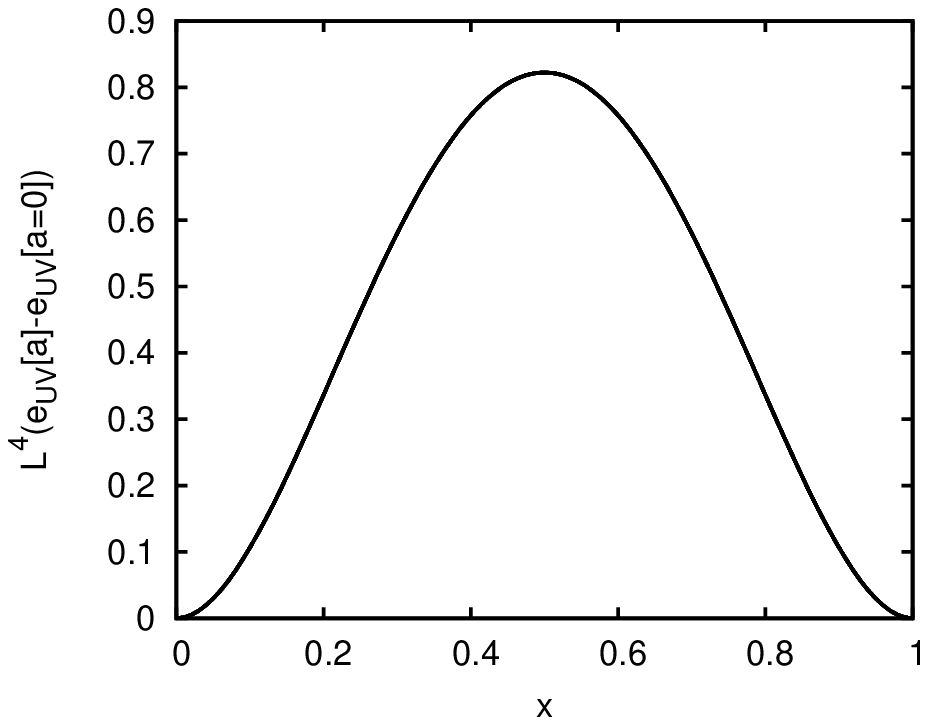}
\caption{The Weiss potential $e_\text{UV}$ multiplied by $L^4$ as a function of the dimensionless field $x=a L /(2 \pi)$.}
\label{fig1-1}
}
\hfil
\centering
\parbox[t]{.45\linewidth}{
\includegraphics[width=\linewidth]{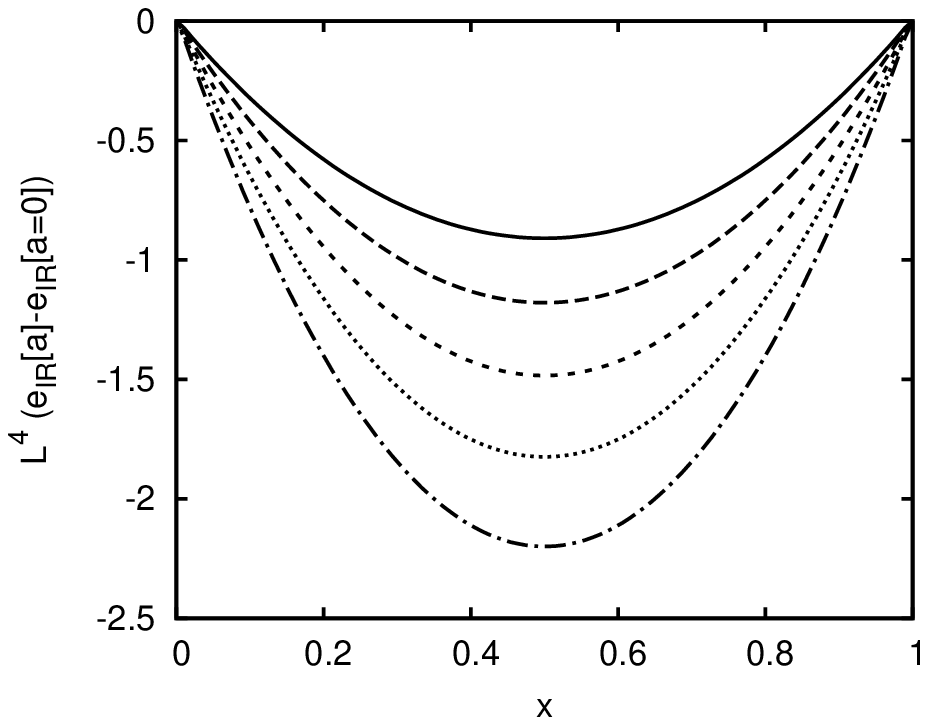}
\caption{The infrared potential $e_\text{IR}$ multiplied by $L^4$ for different $L^{-1}$.}
\label{fig1-2}
}
\end{figure}
Compactifying the $3$-axis to a circle with circumference $L$ and choosing the background field along the compactified dimension $\vec{a}=a \vec{e}_3$ the shifted momentum (\ref{19}) becomes 
\be
\label{24}
\vp^\sigma = \vp_\perp + \lk p_n - \sigma a \rk \ve_3 \, , \quad p_n = 2 \pi n/L \, ,
\ee
where $\vp_\perp$ is the projection of $\vp$ into the $1$-$2$-plane and $p_n$ is the Matsubara frequency. In the Hamiltonian approach the 
effective potential of the constant background field is given by the energy density in the state minimizing $\vev{ H }_a$
under the constraint $\vev{A }_a = a$ \cite{WeinbV2}. Using the gap equation \ref{12} one finds for the energy density per transversal degree of freedom 
 in the present approximation \cite{Reinhardt:2012qe}
\be
\label{25}
e (a, L) = \sum_\sigma \frac{1}{L} \sum^\infty_{n = - \infty} \int \frac{\dd^2 p_\perp}{(2 \pi)^2} \lk \omega \lk \vp^\sigma \rk - 
\chi \lk \vp^\sigma \rk \rk \, .
\ee
By shifting the summation index $n$ one verifies the periodicity
\be
\label{26}
e \lk a + 2 \pi/L , L\rk = e (a, L) \, ,
\ee
which is a necessary property for the effective potential of the confinement order parameter by center symmetry. 
Neglecting $\chi (\vp)$ 
Eq.~(\ref{25}) gives the energy of a non-interacting Bose gas with single-particle energy $\omega (\vp)$. This quasi-particle
picture is a consequence of the Gaussian ansatz (\ref{10}) for the wave functional. The quasi-particle energy $\omega (\vp)$ is,
however, highly non-perturbative, see for example Eq.~(\ref{22}). The curvature $\chi (\vp)$ in Eq.~(\ref{25}) arises from the Faddeev-Popov determinant
in the kinetic part of the Yang--Mills Hamiltonian (\ref{5}).

In certain limiting cases and for $0 \leq a L / 2 \pi \leq 1$ the energy density (\ref{25}) can be calculated analytically. Neglecting $\chi (\vp)$ and assuming the perturbative
expression for the gluon energy $\omega (\vp) = |\vp|$ one finds from (\ref{25}) the Weiss potential originally obtained in \cite{Weiss:1980rj}
\be
\label{27}
e_\text{UV} (a, L) = \frac{4}{3} \frac{\pi^2}{L^4} \lk \frac{a L}{2 \pi} \rk^2 \lk \frac{a L}{2 \pi} - 1 \rk^2 \, 
\ee
shown in Fig.~\ref{fig1-1}.
Neglecting $\chi (\vp)$ and using the infrared expression for the gluon energy
$\omega (\vp) = M^2 / |\vp|$ (see Eq.~(\ref{22})), one obtains \cite{Reinhardt:2012qe}
\be
\label{28}
e_\text{IR} (a, L) = 2 \frac{M^2}{L^2} \left[ \lk \frac{a L}{2 \pi} \rk^2 - \frac{a L}{2 \pi} \right] \, 
\ee
shown in Fig.~\ref{fig1-2}.
This expression drastically differs from the Weiss potential (\ref{27}): While $e_\text{UV} (a, L)$ is minimal for $a = 0$, the minimum of
$e_\text{IR} (a, L)$ occurs at $a = \pi / L$ corresponding to a center symmetric ground state. Accordingly $e_\text{UV} (a, L)$ yields for the 
Polyakov
loop $\langle P \rangle = P [A_0 = 0]  = 1$ while $e_\text{IR} (a, L)$ yields $\langle P \rangle = P [A_0 = \pi / L] = 0$.

\begin{figure}[t]
\centering
\parbox[t]{.45\linewidth}{
\includegraphics[width=\linewidth]{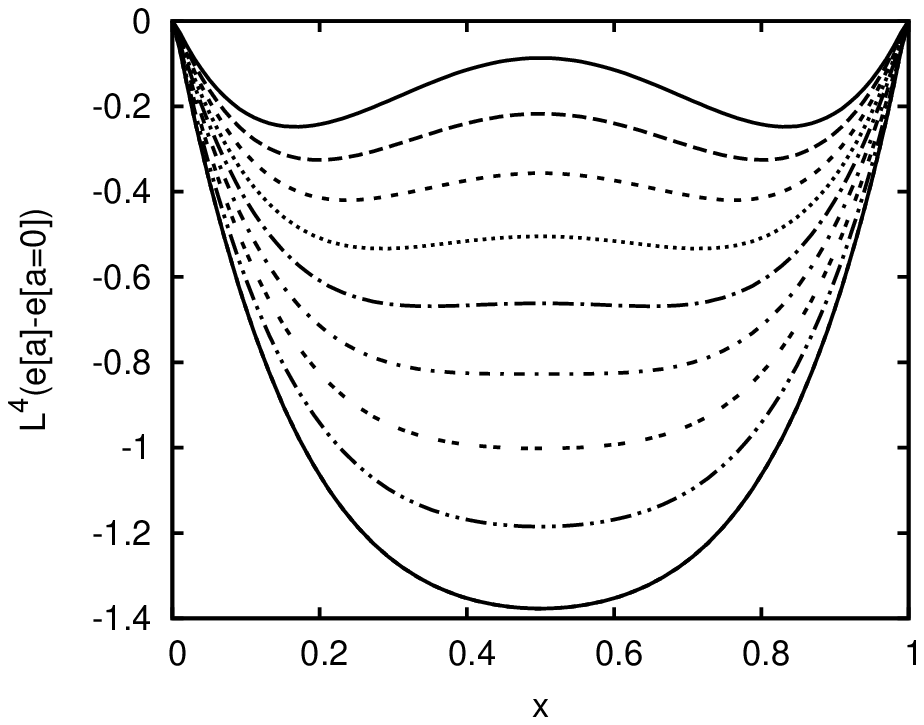}
\caption{The simplified potential (\protect\ref{30}) multiplied by $L^4$ for different temperatures $L^{-1}$. The critical temperature is $T_c \simeq 485$ MeV.}
\label{fig1-3}
}
\hfil
\centering
\parbox[t]{.45\linewidth}{
\includegraphics[width=\linewidth]{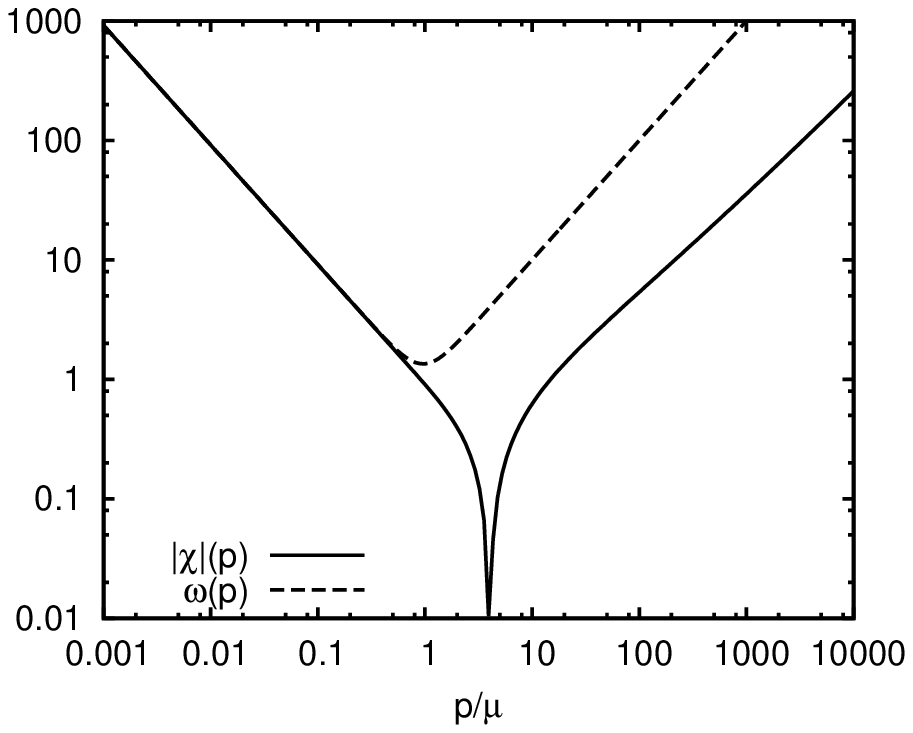}
\caption{The numerical solutions of the gluon energy $\omega$ and the curvature $\chi$ obtained in \cite{HefReiCam12} .}
\label{fig2-1}
}
\end{figure}
The deconfinement phase transition results from the interplay between the confining infrared potential, Eq.~(\ref{28}) and the deconfining 
UV-potential, Eq.~(\ref{27}). 
To illustrate this let us approximate the gluon energy $\omega(\vp)$
(\ref{22}) by 
\be
\label{29}
\omega (\vp) \approx |\vp| + M^2/|\vp| \, .
\ee
This expression agrees with the Gribov formula (\ref{22}) in both, the IR and UV but deviates from it in the mid-momentum regime, which
influences the deconfinement phase transition. With $\omega (\vp)$ given by Eq.~(\ref{29}) and with $\chi (\vp) = 0$
the energy density (\ref{25}) becomes
\begin{gather}
\label{30}
e (a, L)  = e_\text{IR} (a, L) + e_\text{UV} (a, L) = \frac{4}{3} \frac{\pi^2}{L^4} f \lk \frac{a L}{2 \pi} \rk\,, \\
f (x)  =  x^2 (x -1)^2 + c x (x - 1) , \, \quad c = \frac{3 M^2 L^2}{2 \pi^2} \nonumber\,.
\end{gather}
For small temperatures $L^{- 1}$, $e_\text{IR} (a, L)$ dominates and the system is in the confined phase. As $L^{- 1}$ increases the center symmetric
minimum at $x = 1/2$ eventually turns into a maximum and the system undergoes the deconfinement phase transition, see Fig.~\ref{fig1-3}. In the deconfined phase 
$f (x)$ has two degenerate minima and, starting in the deconfined phase, the phase transition occurs when the three roots of $f' (x)$ degenerate.
This occurs for $c = 1/2$, i.e. for a critical temperature
\be
\label{31}
T_c = L^{- 1} = \sqrt{3} M / \pi \, .
\ee
With the lattice result $M = 880$ MeV this corresponds to a critical temperature of $T_c \simeq 485$ MeV, which is much too high. This value is only slightly reduced to $T_c = 432$ MeV when the correct Gribov formula (\ref{22}) is used instead of the
approximation (\ref{27}). The reason why the transition temperature comes out too high is the neglect of the ghost loop $\chi (p) = 0$.
This can be seen from Fig.~\ref{fig2-1}, where the gluon energy $\omega (p)$ and the curvature $\chi (p)$ obtained from the variational calculation
in Coulomb gauge are shown. In the deep infrared the gluon energy $\omega (p)$ and the curvature $\chi (p)$ agree, while in the ultraviolet 
$\chi (p)$ has the opposite sign of $\omega (p)$ and is only suppressed by a log compared to $\omega (p)$. Therefore, neglecting the ghost
loop enhances the infrared contribution to the potential, which favors confinement, while at the same time it reduces the ultraviolet 
contribution, which favors deconfinement. Both effects add coherently and push the deconfinement transition to a higher temperature. 
\begin{figure}[t]
\centering
\parbox[t]{.45\linewidth}{
\includegraphics[width=\linewidth]{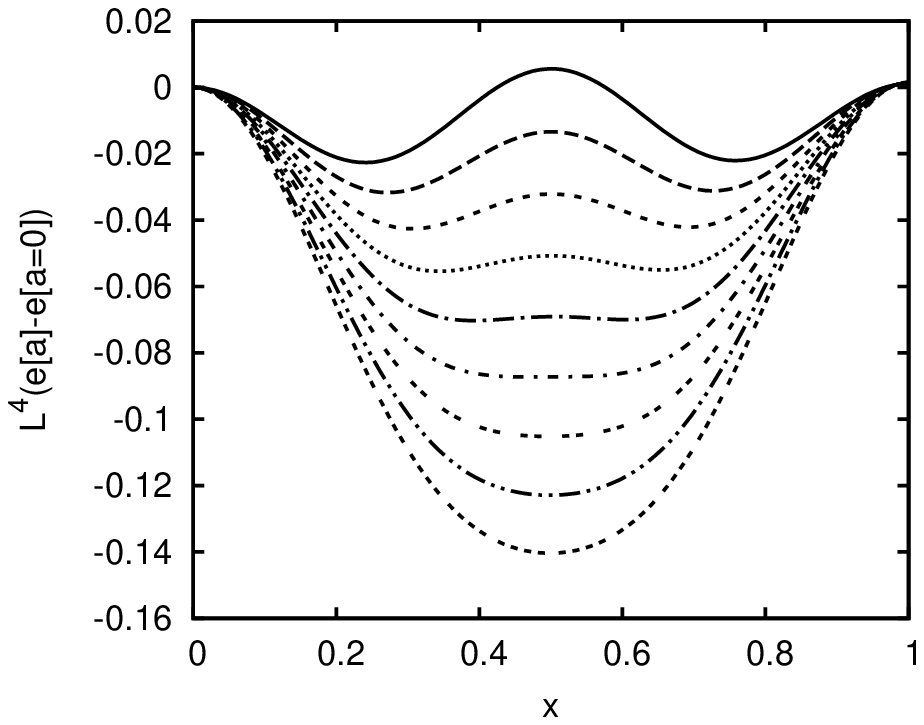}
\caption{The numerically evaluated effective potential. The critical temperature is $T_c \simeq 269$ MeV.}
\label{fig3}
}
\hfil
\centering
\parbox[t]{.45\linewidth}{
\includegraphics[width=\linewidth]{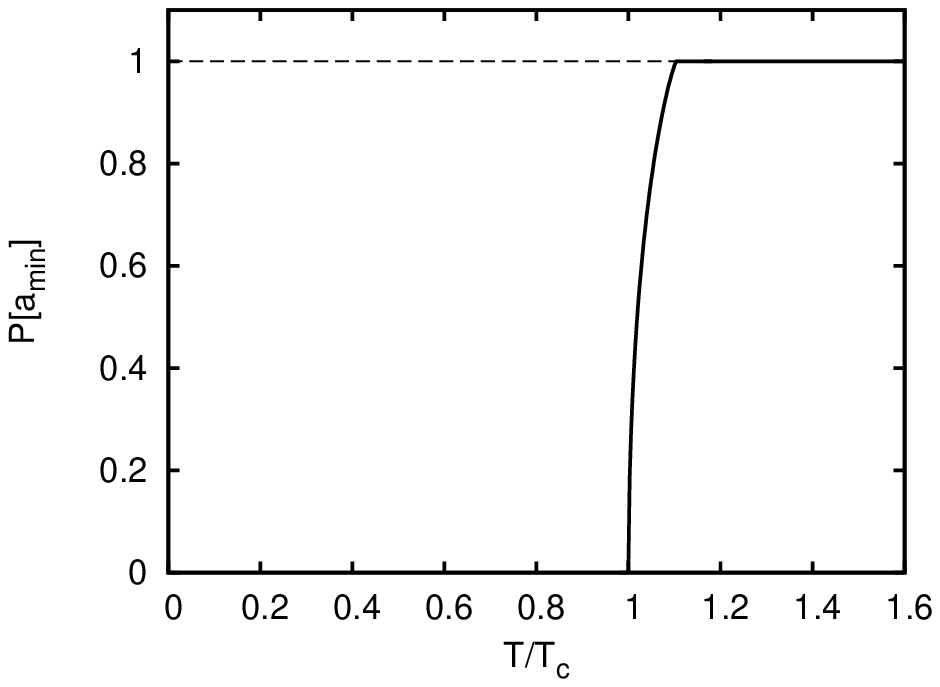}
\caption{The Polyakov loop $\vev{P[a]}$  evaluated at the minimum $a = a_\text{min}$ of the full effective potential (figure (a)) as a function of $T/T_c$.}
\label{fig4}
}
\end{figure}

In a full numerical evaluation of the effective potential (\ref{25}) using for $\omega (\vp)$
and $\chi (\vp)$ the numerical solution of the variational approach in Coulomb gauge obtained in Ref.~\cite{HefReiCam12}, one finds the effective
potential shown in Fig.~\ref{fig3}. From this potential one extracts a critical temperature for the deconfinement phase transition of $T_c \simeq 269$ MeV,
which is close to the lattice predictions of $T_c = 290$ MeV. This value is also close to the range of critical temperatures $T_c = 275 \ldots 290$ MeV obtained in Ref.~\cite{HefReiCam12}
from the grand canonical ensemble of Yang--Mills theory in Coulomb gauge. Let us also mention that if one uses for $\omega (\vp)$
the Gribov formula (\ref{22}) and in accord with the gap equation (\ref{12}) for $\chi (\vp)$ its infrared expression (\ref{23}) one finds a 
critical temperature of $T_c \simeq 267$ MeV, which is only slightly smaller than the value obtained with the full numerical solution for $\omega (\vp)$ and
$\chi (\vp)$. This shows that it is the infrared part of the curvature (neglected in Eq.~(\ref{30})), which is crucial for the critical
temperature. In view of the ghost dominance in the IR this is not surprising. Fig.~\ref{fig4} shows the Polyakov loop $P[a]$ calculated from the minimum $a_\text{min}$ of the potential (\ref{25}). At the phase-transition $P[a_\text{min}]$ rapidly changes from $P=0$ to $P=1$, which is typical for a second order phase transition.

In the present approach the deconfinement phase transition is entirely determined by the zero-temperature propagators, which are defined as 
vacuum expectation values. Consequently, the finite-temperature behavior of the theory and, in particular, the dynamics of the deconfinement
phase transition must be fully encoded in the vacuum wave functional. The results obtained above are encouraging for an 
extension of the present approach to full QCD at finite temperature and baryon density.

\end{document}